\documentclass[10 pt]{iopart}
\usepackage{graphicx}

\expandafter\let\csname equation*\endcsname\relax
\expandafter\let\csname endequation*\endcsname\relax
\usepackage{amsmath, color, ulem}

\newcommand{\ve}[1]{{\bf #1}}

\newcommand{\mus}[0]{\mu_\mathrm{s}}

\usepackage{upgreek}
\usepackage{siunitx}
\usepackage{cite}

\definecolor{green}{rgb}{0.2,0.4,0.0}

\usepackage[colorlinks=true]{hyperref}

\begin{document}


\title{Spin transport across antiferromagnets induced by the spin Seebeck effect}
\author{Joel Cramer$^{1,2}$, Ulrike Ritzmann$^{3,4}$, Bo-Wen Dong$^{1,2,5}$, Samridh Jaiswal$^{1,6}$, Zhiyong Qiu$^{7}$, Eiji Saitoh$^{7,8,9,10}$, Ulrich Nowak$^{\ast,4}$, Mathias Kl\"aui$^{\star,1,2}$}

\address{$^1$ Institute of Physics, Johannes Gutenberg-University Mainz, 55099 Mainz, Germany}
\address{$^2$ Graduate School of Excellence Materials Science in Mainz (MAINZ), Mainz, 55128 Mainz, Germany }
\address{$^3$ Department of Physics and Astronomy, Uppsala University, 751 05 Uppsala, Sweden}
\address{$^4$ Department of Physics, University of Konstanz, 78457 Konstanz, Germany}
\address{$^5$ School of Materials Science and Engineering, University of Science and Technology Beijing, 100083 Beijing, People's Republic of China}
\address{$^6$ Singulus Technologies AG, 63796 Kahl am Main, Germany}
\address{$^7$ Advanced Institute for Materials Research, Tohoku University, Sendai 980-8577, Japan}
\address{$^8$ Institute for Materials Research, Tohoku University, Sendai 980-8577, Japan}
\address{$^9$ Center for Spintronics Research Network, Tohoku University, Sendai 980-8577, Japan}
\address{$^{10}$ Advanced Science Research Center, Japan Atomic Energy Agency, Tokai 319-1195, Japan}
\eads{$^{\ast}$\mailto{ulrich.nowak@uni-konstanz.de}, $^{\star}$\mailto{klaeui@uni-mainz.de}}

\begin{abstract}
For prospective spintronics devices based on the propagation of pure spin currents, antiferromagnets are an interesting class of materials that potentially entail a number of advantages as compared to ferromagnets.
Here, we present a detailed theoretical study of magnonic spin current transport in ferromagnetic-antiferromagnetic multilayers by using atomistic spin dynamics simulations. The relevant length scales of magnonic spin transport in antiferromagnets are determined.
We demonstrate the transfer of angular momentum from a ferromagnet into an antiferromagnet due to the excitation of only one magnon branch in the antiferromagnet.
As an experimental system, we ascertain the transport across an antiferromagnet in YIG$|$Ir\textsubscript{20}Mn\textsubscript{80}$|$Pt heterostructures.
We determine the spin transport signals for spin currents generated in the YIG by the spin Seebeck effect and compare to measurements of the spin Hall magnetoresistance in the heterostructure stack.
By means of temperature-dependent and thickness-dependent measurements, we deduce conclusions on the spin transport mechanism across IrMn and furthermore correlate it to its paramagnetic-antiferromagnetic phase transition.
\end{abstract}


\maketitle
\ioptwocol

\section{Introduction}

For the development of next-generation, energy-efficient spintronic devices for information transmission, processing, and storage, the investigation of pure spin currents has attracted great interest during recent years.
In contrast to spin-polarized charge currents, with a broad spectrum of applications in current spintronics schemes (e.g. spin-transfer-torque operated magnetic tunnel junctions \cite{ikeda2010perpendicular}), pure spin currents exclusively transfer angular momentum and have no net charge flow.
While in normal metals that exhibit the spin Hall effect \cite{Sinova2015} pure spin currents are realized by charge currents of opposite spin-polarization flowing in opposite directions, magnetically ordered systems provide a further spin transport channel via magnonic (spin wave) excitations with no moving charges \cite{kruglyak2010magnonics}.
Aside from information transfer and data handling, pure spin currents have furthermore proven as a useful tool to investigate magnetic material properties.
Spin Hall magnetoresistance (SMR) \cite{Nakayama2013} measurements, for instance, allow to probe the orientation of magnetic sublattice moments in complex magnetic oxides \cite{Ganzhorn2016,dong2017spin}, which are otherwise not accessible using common characterization methods, e.g. SQUID magnetometry.


With respect to magnonic spin current propagation, insulating ferromagnets (FM) pose an interesting medium and therefore caught notable renewed attention in recent years.
As compared to metallic systems, insulators prevent spin transfer mediated by charge motion and consequently do not exhibit Joule heating losses within the insulator.
The most prominent representative of this material class is the yttrium iron garnet Y$_3$Fe$_5$O$_{12}$ (YIG) \cite{cherepanov1993saga}, since it reveals excellent insulating properties and extremely low Gilbert damping \cite{chang2014nanometer}.
In single crystalline YIG, magnon propagation lengths in the range of several micrometer have been reported \cite{Cornelissen2015,Goennenwein2015,Kehlberger2015,Guo2016}.
More recently, however, due to potential advantages over ferromagnets antiferromagnets (AFM) have gained increased interest considering spintronics applications \cite{jungwirth2016antiferromagnetic}.
In AFMs, neighboring magnetic moments are ordered alternatingly, such that the macroscopic moment $M$ of the solid vanishes.
As a result, adjacent AFM devices do not exhibit mutual interaction due to the lack of stray fields and furthermore are insensitive to external magnetic field perturbations.

It has been shown that insulating AFMs are able to exhibit thermal magnon currents induced by the spin Seebeck effect (SSE) \cite{Bauer2012,Uchida2014,Kehlberger2015} when driven into the spin-flop state \cite{Seki2015,Wu2016,Rezende2016a}.
Magnon propagation across AFM thin films has been investigated in FM/AFM/HM heterostructures both experimentally \cite{Hahn2014,Wang2014,Wang2015,Qiu2016,Lin2016,Prakash2016} and theoretically  \cite{Rezende2016,Chen2016,Khymyn2016}.
Since the excitation frequency of antiferromagnetic magnons usually lies in the range of several THz, they cannot be excited by optical or current electrical methods.
Therefore, spin currents are generated in the FM layer, pumped into the AFM, and eventually detected in the HM by means of the inverse spin Hall effect (ISHE).
The change of the ISHE voltage signal when measured as a function of AFM thickness or temperature eventually then allows one to infer information on the AFM magnon propagation properties.

Here, we put forward an analytical model describing the details of magnon propagation in antiferromagnets. We demonstrate an exponential spatial decay of AFM magnons in insulators, which is in line with experimental observations. Despite the high speed of antiferromagnetic magnons, their range is limited due to a very short life time. Though the propagation length reveals a clear maximum just above the energy gap, it is significantly smaller as compared to ferromagnetic systems.
Our analytical work is well in agreement with the results obtained for atomistic spin dynamics simulations. Moreover, we present angular momentum transfer due to magnon propagation from a FM into an AFM. We identify two different regimes: Below the frequency gap, evanescent modes with a very strong spatial decay are excited within the AFM. Above the frequency gap, antiferromagnetic magnons are excited that propagate on a longer range within the AFM.

On the experimental side, we investigate spin current transmission across the  metallic AFM Ir\textsubscript{20}Mn\textsubscript{80} (IrMn) in YIG/IrMn/Pt trilayers to identify potentially dominant spin transmission channels (electronic vs. magnonic).
We perform both SSE and SMR measurements and compare the temperature- and thickness-dependent signal amplitudes obtained to examine whether genuine spin transport across IrMn or interface exchange coupling phenomena are observed.
It was shown before that the thickness-dependent antiferromagnetic-paramagnetic phase transition of IrMn thin films can be probed by means of temperature-dependent ferromagnetic resonance spin pumping measurements \cite{Frangou2016}.
In trilayers of Ni\textsubscript{81}Fe\textsubscript{19}/Cu/IrMn, the Gilbert damping constant $\alpha$ of Ni\textsubscript{81}Fe\textsubscript{19} exhibits an enhancement near $T$\textsubscript{N\'{e}el}, revealing increased spin sink properties of the IrMn layer for the pumped spin current due to spin fluctuations.
As similar observations were made for systems including insulating AFMs \cite{Qiu2016,Lin2016,Prakash2016}, this implies a significant coupling of the spin current to the antiferromagnetic ordering parameter in IrMn.
Consequently, this method allows one to indirectly gain insight into the magnetic properties of IrMn.
While no direct information about spin propagation was previously obtained, we here compare different layer stacks to identify the spin transport contribution to the signal.

\section{Analytical model of magnon propagation in ferromagnets and antiferromagnets} 
We start the development of the theoretical model by discussing spin transport in FMs and AFMs individually and the length scales involved. For that purpose, we consider a simple cubic lattice with lattice constant $a$. In the Hamiltonian, we include exchange interaction of nearest neighbors with exchange constant $J$ and an anisotropy leading to an easy axis in $x$-direction with anisotropy constant $d_x$. The Hamiltonian is then given by
\begin{align}
\mathcal{H}=\sum_{\langle ij \rangle}{J_{ij}\ve{S}_i\ve{S}_j}+\sum_{i}{d_x\left(S_{i}^x\right)^2}\;\mbox{.}
\end{align}
We perform atomistic spin dynamics simulations \cite{Nowak2007} as well as analytical calculations based on the Landau-Lifshitz-Gilbert (LLG) equation,
\begin{align}
\dot{\ve{S}}=-\frac{\gamma}{\mus(1+\alpha^2)}\ve{S}\times\ve{H}-\frac{\gamma\alpha}{\mus(1+\alpha^2)}\ve{S}\times(\ve{S}\times\ve{H})\mbox{.}
\end{align}
This equation of motion describes the precession of normalized magnetic moments $\ve{S}$ around their effective field $\ve{H}=-\partial \mathcal{H}/\partial \ve{S}$ and relaxation depending on the damping constant $\alpha$. $\gamma$ denotes the gyromagnetic ratio and $\mus$ is the magnetic moment. 

In Ref. \cite{Ritzmann2014}, the propagation length of magnons was investigated for FM systems. By linearizing the LLG equation and assuming $\ve{S}\approx\ve{e}_x$, the coupled equations of motions were solved.
The imaginary part of the eigenvalue defines the magnon frequency
\begin{align}
 \hbar\omega_\mathrm{FM}=2d_x+J\sum_{\theta}{(1-\cos(q_\theta a))}\;\mbox{,}
\end{align}
where $\theta$ denotes the cartesian components. The real part describes the lifetime $\tau=1/(\alpha\omega)$. A magnon accumulation was defined as the transferred magnetic moment that scales with $\Delta m\approx\sum_{\ve{q}}1/2A_{\ve{q}}^2$, where $A_\ve{q}$ is the spin wave amplitude. Considering a spin wave propagating only in $z$-direction, $\ve{q}=q_z\ve{e}_z$, the propagation length was defined as the decay of the magnon accumulation $\Delta m$ and was obtained via  the lifetime $\tau$ and the group velocity $v_z=\partial\omega/\partial q_z$. The result for the propagation length was
\begin{align}
\xi_\mathrm{FM}(\omega)=\frac{\tau v_z}{2}=\frac{J a}{2\alpha\hbar\omega}\sqrt{1-\Big(1-\frac{\hbar\omega-2d_x}{J} \Big)^2}\;\mbox{.}
\end{align}
The propagation length has a maximum close to the frequency gap and decays with increasing frequency. For low damping and low anisotropies, the propagation length of low frequency magnons is in the range of up to a few $\mu$m. These results explain a saturation effect of the SSE in YIG and a suppression effect due to large external magnetic fields \cite{Kehlberger2015, Ritzmann2015a}.

Here, we now develop the analogous model for AFMs. We consider a similar system and choose $J<0$. This system consists of two sublattices A and B. To describe magnon excitations, we linearize the LLG equation for each sublattice and assume $S_{i,A}^x\approx1$ and $S_{i,B}^x\approx-1$, as well as a small damping constant $\alpha\ll1$.

The considered AFM has two magnon branches. A magnon describes a collective precession of magnetic moments in both sublattices, but with unequal amplitudes. The ratio of the amplitudes of the two sublattices is wave-vector dependent and it is reversed for the two magnon branches. Therefore, magnons of opposite branches carry opposite angular momentum. Moreover, magnetic moments precess either all clockwise or counterclockwise within the two different magnon branches. In the absence of an external magnetic field,  the magnon branches are degenerate and their dispersion relation is given by
\begin{align}
 \hbar\omega_\mathrm{AFM}=\sqrt{\big(2d_x+6|J|\big)^2-4J^2\big(\sum_{\theta}{\cos(q_\theta a)}\big)^2}\;\mbox{.}
\end{align}

  \begin{figure}[t!]
   	\centering
   	\includegraphics[trim=1.cm 0 0 0, clip,width=0.4\textwidth]{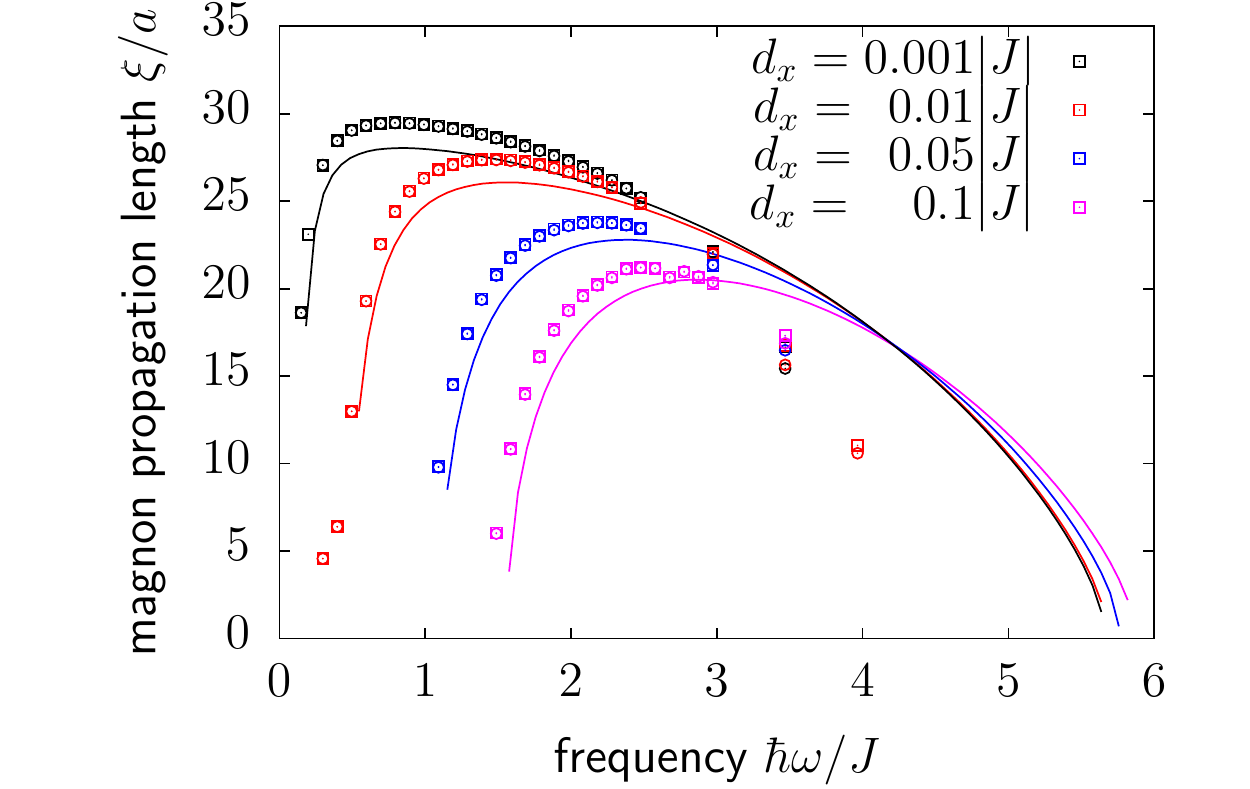}
   	\caption{Frequency dependent magnon propagation length in an AFM for various anisotropy constants $d_x$ and a damping constant of $\alpha=0.01$. Numerical data, depicted as data points, are in agreement with the analytical model, which is shown as corresponding continuous lines.}
   	\label{xi}
  \end{figure}
In contrast to FMs, AFMs have a large frequency gap of $\hbar\omega_0 \approx \sqrt{24d_x |J| }$.
Due to degeneracy, magnons from both branches are excited thermally with equal probablity and no magnetitization occurs at constant temperatures.
The total magnetization is also compensated in linear temperature gradients.
It was shown that around a temperature step no net spin transfer occurs in AFMs, although a magnon current appears \cite{Ritzmann2017}.

Despite the fact that in an isolated AFM no net spin current occurs, the length scale of magnon propagation in AFMs is interesting to study. Thermally induced magnons do not transfer angular momentum, but they still transfer heat and are the origin of thermally driven domain wall motion in AFMs \cite{Selzer2016, Tveten2014}. Moreover, external magnetic fields lift the degeneracy. It has been shown, that thermally activated spin currents appear due to the SSE \cite{Wu2016, Rezende2016a}.  

The lifetime of AFM magnons is given by the real part of the eigenvalue and one obtains
\begin{align}
 \tau=\frac{\hbar}{(2d_x+6|J|)\cdot\alpha}\;\mbox{.}
\end{align}
The resulting lifetime is shorter than in FMs and independent of the magnon frequency. We obtain for the frequency-dependent magnon propagation length
\begin{align}
  \label{xi_afm}
   \xi(\omega)=&\frac{a|J|\sqrt{H_0
   ^2-(\hbar\omega)^2}}{\alpha H_0\hbar\omega}\sqrt{1-\Big(\frac{\sqrt{H_0^2-(\hbar\omega)^2 }}{2|J|}-2\Big)^2}\;\mbox{,}
  \end{align}
  where we use the abbreviation $H_0=2d_x+6|J|$.

We simulate the decay of magnons in an AFM with $8\times8\times512$ magnetic moments. To excite monochromatic spin waves with a group velocity only in $z$-direction, we attach an additional layer in the $x$-$y$-plane, in which all magnetic moments precess homogeneously with frequency $\omega$. The magnetic moments of the two sublattices are aligned in oppposite directions and their precession has a phase shift of 180 degrees.  Due to exchange interaction, this layer couples to the system and monochromatic spin waves enter. By fitting the exponential decay of the spin wave amplitudes, we calculate their propagation length. 

The results from the analytical formula as well as from numerical simulations are shown in Fig. \ref{xi}. The propagation length increases strongly just above the frequency gap $\omega_0$ until a maximum value is reached and then decreases with further increasing frequency. The maximum values are much shorter than in FMs. Despite the higher velocity for magnons at a frequency close to gap, their range is still small due to their short lifetime. As shown in the figure, the analytical formula describes the general behavior of the propagation length. However, for high frequencies deviations between analytical calculation and numerical simulation appear due to the limited cross section in the simulations. Note that in contrast to FMs, the dispersion relation of AFMs depends on the spatial dimension of the lattice.

    \begin{figure}[t]
   \centering
   \includegraphics[width=0.44\textwidth]{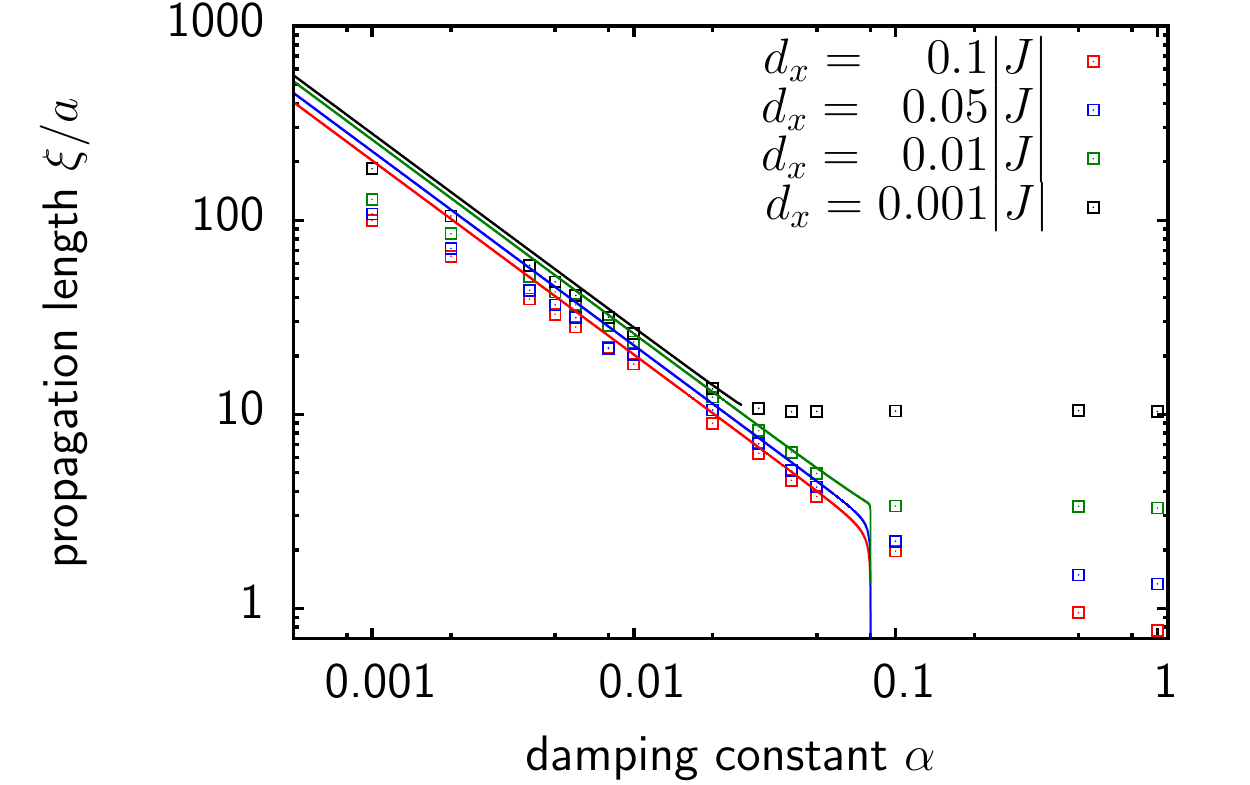}
   \caption{Magnon propagation length $\xi$ as a function of the damping constant $\alpha$ for different anisotropy constants $d_x$. The numerical data are shown as data points and the continuous lines represent the maximum value of the analytical one-dimensional model.}
   \label{afm2}
  \end{figure} 

Similar to our previous studies on FMs \cite{Ritzmann2014}, we study the length scale of thermally triggered magnon propagation using a temperature step to excite the magnons. We simulate $8\times8\times512$ magnetic moments and apply a temperature step along the $z$-axis from $k_\mathrm{B}T_1=0.1|J|$ to $k_\mathrm BT_2=0$. We fit the decay of the magnon accumulation in both sublattices and compare the resulting length scale with the maximum propagation length from equation \ref{xi_afm}. Figure \ref{afm2} shows the results from numerical simulation as well as from the analytical model. For high damping values, the analytical formula deviates since we neglected $\alpha^2$- terms in the derivation. But both methods give similar results for low damping values. In contrast to low frequency magnons in FMs, which can propagate over several $\mu$m, the AFM magnons have a much shorter range in the nm-regime.

\section{Magnon transfer in ferromagnet-antiferromagnet-heterostructures} 
To compare to experimental work, we study the excitation of spin waves in hetero- structures consisting of a FM and an AFM layer. We excite a monochromatic spin wave in the FM and study the transfer of angular momentum into the AFM. We perform simulations with a FM system with $8\times8\times256$ magnetic moments and additionally an AFM layer of the same size attached to it. For simplification, we use a layered AFM by considering antiferromagnetic exchange interaction only in $z$-direction, $J_\mathrm{FM}=J_\mathrm{AFM}^x=J_\mathrm{AFM}^y=-J_\mathrm{AFM}^z$. The exchange interaction at the interface is given by $J_\mathrm{IF}=-J_\mathrm{FM}$. The monochromatic spin wave is excited by a homogenous precession of the magnetic moments with a given frequency $\omega$ at the 0th layer of the FM. 

\begin{figure}[t]
   \centering
   \includegraphics[trim=1.cm 0 0 0, clip,width=0.4\textwidth]{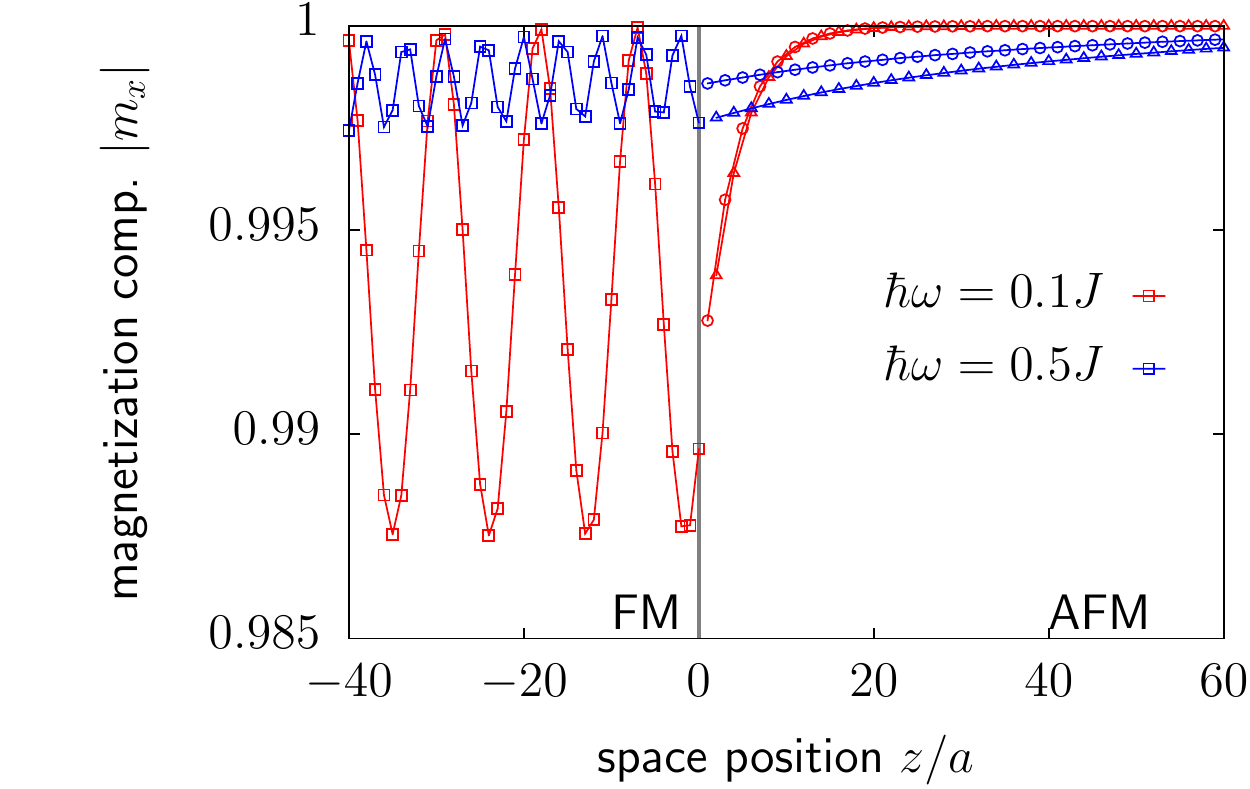}\caption{\label{bilayer} Absolute value of the $x$-component of the magnetization for spin wave propagation from a FM ($z<0$) to AFM ($z>0$) layer for an evanescent mode ($\hbar\omega=0.1J$) and a normal mode ($\hbar\omega=0.5J$). The dots (triangles) in the AFM-regime show $m_x$ for sublattice A (B). } 
   \label{rotevan}
    \end{figure}%
Dependent on the frequency of the spin wave, two different regimes for the spin wave propagation within the AFM appear. For frequencies below the gap of the dispersion relation of the AFM, the signal decays exponentially with distance to the interface. These are evanescent modes  \cite{Khymyn2016}. Spin waves with frequencies above the gap excite a spin wave of the same frequency within the antiferromagnet. Note that in this quasi one-dimensional AFM, the dispersion relation is given by 

\begin{align}
   \hbar\omega=\sqrt{\big(2d_x+2|J|\big)^2-4J^2\big(\cos(q_z a)\big)^2}\;\mbox{.}
   \label{afmgap}
\end{align}
The frequency gap in this case is $\hbar\omega_0 \approx \sqrt{8 d_x |J|}$.

In Fig. \ref{bilayer}, we show the $x$-component (easy axis) of the magnetization for two different examples. The red curves show an evanescent mode where no precession of the $y$- and $z$- components of the magnetization  in the AFM is observed and the signal disappears on a very short length scale. The blue curves represent a normal mode in the AFM, where the spin wave propagates within the AFM with the same frequency as in the FM. The $y$- and $z$- components  of the magnetization of the single sublattices show precession due to AFM spin wave propagation, whereas the $x$-component of the magnetization in both sublattices decays exponentially within the magnon propagation length.

The orientation of the magnetization of the FM determines the sense of the rotation of the magnetic moments as well as the transferred angular momentum in the FM. Therefore, only one of the two magnon branches is excited and due to the different amplitudes of the two sublattices, angular momentum is transferred. The oscilattion of the $x$-component within the FM layer illustrates interference of the incoming spin wave with a strongly reflected wave at the interface and only a small ratio of the signal is transferred in both cases into the AFM. Note that both spin waves in the ferromagnet have been excited with the same initial amplitude at $z=-256a$. The higher frequency has a much shorter propagation length in the FM and, therefore, its amplitude at the interface is significantly smaller. Nevertheless, with larger distances to the interface, the normal AFM magnon causes a larger signal than the evanescent mode. The chosen frequency is close to the gap and the propagation length is several nm.

Here, we demonstrate the propagation of spin waves for a single monochromatic wave. For temperature gradients inducing the SSE a broad frequency spectra would be excited in the ferromagnetic layer. Due to the larger propagation length at low frequencies within the FM, these frequencies play an important role in the SSE in YIG \cite{Ritzmann2015a}. Due to the high frequency gap of antiferromagnets, mainly evanescent modes should be excited. The transferred spin current should decay exponentially within distances in the range of a few nm.

\section{Experimental investigation of spin current transmission across a metallic antiferromagnet}

Having established the theory of spin transport in and across AFMs using pure magnonic spin currents, we next investigate spin transport experimentally in a combination of ferromagnetic, antiferromagnetic and heavy metal layers. 

To begin with, let us compare the results of the theory to experimental findings for systems including insulating AFMs, where the spin current can only be carried by magnons.
The extensive literature \cite{Hahn2014,Wang2014,Wang2015,Qiu2016,Lin2016,Prakash2016} shows that indeed an exponential decay of the signal is found with increasing thickness of the AFM.
So qualitatively, in these systems the theoretical description seems to hold and is apt to describe the spin transport mechanism.
As a next step, we probe here experimentally the spin current transport in conducting AFMs.
In systems including the latter, the spin current can be transported by magnons as described above, but additionally also by charge-based spin currents.
To check if charge-mediated transport of spin information occurs in addition to the magnonic spin currents described above, we performed temperature-dependent spin transmission experiments in a stack including the metallic AFM Ir\textsubscript{20}Mn\textsubscript{80} (IrMn) using YIG/IrMn/Pt trilayers.

In the experiment, spin currents are either triggered by the spin Seebeck effect \cite{Bauer2012,Uchida2014,Kehlberger2015} or via the spin Hall effect using spin Hall magetoresistance measurements \cite{Nakayama2013}.
As a first difference to insulating AFMs, one has to take into account the fact that in addition to Pt, which is widely used as a model material for ISHE based experiments, IrMn itself as well exhibits a spin Hall effect \cite{Mendes2014}.
Therefore, in order to understand this more complex system, one needs to study not just the trilayer YIG/IrMn/Pt but also the individual combinations YIG/IrMn and YIG/Pt.
Initially, single crystalline YIG is grown epitaxially on (111)-oriented Gd\textsubscript{3}Ga\textsubscript{5}O\textsubscript{12} (GGG) substrates by liquid-phase-epitaxy with a film thickness of \SI{5}{\micro\meter}.
Onto GGG/YIG samples of size \SI{2 x 6 x 0.5}{\milli\meter}, IrMn/Pt bilayers with varying IrMn thickness but constant Pt thickness ($d_{\mathrm{IrMn}} = 0.8, 1.3 \, \si{\nano\meter}$, $d_{\mathrm{Pt}} = \SI{5}{\nano\meter}$) are deposited via magnetron sputtering.
Furthermore, YIG/Pt($d_{\mathrm{Pt}} = \SI{5}{\nano\meter}$) and YIG/IrMn ($d_{\mathrm{IrMn}} = \SI{1.3}{\nano\meter}$) reference samples are fabricated for comparison.

The temperature-dependent SSE measurements are performed in a cryostat with a variable temperature insert ($\SI{5}{\kelvin} \leq T \leq \SI{300}{\kelvin}$), employing the conventional longitudinal configuration \cite{Uchida2016,Kehlberger2015}.
By sandwiching the samples in between a top resistive heater and a bottom temperature sensor, an out-of-plane (\textit{z} direction) temperature gradient is generated, which induces the thermal spin current in the YIG layer.
Base temperature and temperature gradient are determined via the resistance change of heater and sensor.
An external magnetic field $H$ is applied in-plane along the sample short edge (\textit{y} direction), such that a detectable ISHE voltage drop in the long axis of the sample (\textit{x} direction) appears.
The SSE voltage $V$\textsubscript{SSE} is extracted from the difference between the ISHE voltages obtained for positive and negative magnetic field divided by 2.
To account for the different film resistivities, the SSE current  $I_{\mathrm{SSE}} = V_{\mathrm{SSE}} /R $ is considered in the following.

The temperature-dependent SMR measurements are carried out in a superconducting vector cryostat that allows to align the magnetic field in all directions.
The SMR ratio is extracted from angular-dependent resistance measurements, in which the magnetic field $H$ is rotated in the \textit{yz}-plane and a $\sin ^2 \varphi_{yz}$ resistance change [low (high) resistance for $H$ in-plane (out-of-plane)] is observed.
To ensure that the magnetization follows the applied field direction, the field strength is fixed to a value of $\mu_0 H = \SI{0.8}{\tesla}$, which is much larger than the coercivity of the YIG.

In the following, we start by describing the experimentally determined spin signals as a function of temperature.
Then, in a second step we discuss the results of the different measurements and the implications for the spin transport that we can deduce.

\begin{figure}[!bt]
	\centering
	\includegraphics[width=7.53 cm]{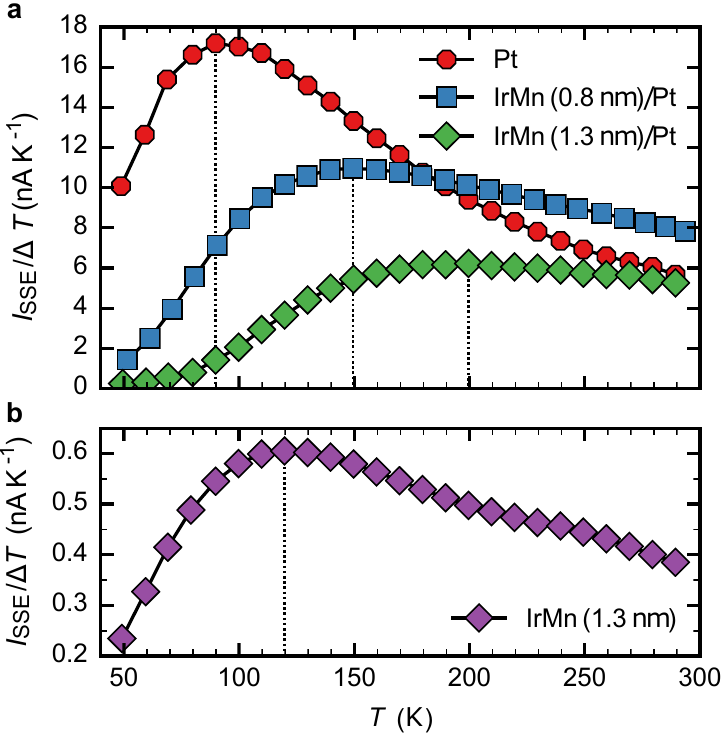}
	\caption{Detected spin Seebeck current as a function of temperature for (a) YIG/Pt or YIG/IrMn/Pt and (b) YIG/IrMn bi- and tri-layers.}
	\label{fig:sse}
\end{figure}

First, we show in Fig. \ref{fig:sse} the measured SSE current amplitude divided by the temperature difference between sample top and bottom as a function of temperature for the stacks investigated.
For enhanced readability, the data obtained for the samples with and without a Pt top layer are presented separately in Fig. \ref{fig:sse}a and Fig. \ref{fig:sse}b.
The YIG/Pt only sample (red circles) exhibits a clear signal maximum near $T = \SI{90}{\kelvin}$, whereas broad, flat maxima are observed at different temperatures for the samples with the additional IrMn interlayer.
For the samples with IrMn layers, the detected SSE signal amplitudes become significantly suppressed at low temperatures below the maxima [$T_{\mathrm{crit}}(d_{\mathrm{IrMn}} = \SI{0.8}{\nano\meter}) \approx \SI{150}{\kelvin}$, $T_{\mathrm{crit}}(d_{\mathrm{IrMn}} = \SI{1.3}{\nano\meter}) \approx \SI{200}{\kelvin}$].
We find at low temperatures, where the IrMn orders antiferromagnetically, that the insertion of IrMn generally yields a thickness-dependent signal reduction, which is in line with the theory described above.
However, at higher temperatures ($T \geq \SI{200}{\kelvin}$), where the IrMn is likely in the paramagnetic phase, a larger $I_{\mathrm{sse}}/\Delta T$ amplitude is observed for YIG/IrMn (\SI{0.8}{\nano\meter})/Pt as compared to the YIG/Pt sample.
This behavior clearly goes beyond the theoretical description put forward above, since there only the AFM phase is considered.
Possible origins of this behavior include an enhanced effective spin-mixing conductance of the YIG/IrMn interface as compared to the YIG/Pt interface \cite{Kikuchi2015,yuasa2017spin}.
While of interest, this aspect is however not the focus of this work and further studies are necessary to understand this, which go beyond the scope of the current work.
Finally, comparing the samples with and without Pt capping layers, we see that the temperature dependence of $I_{\mathrm{sse}}$ for YIG/IrMn (\SI{1.3}{\nano\meter}) in Fig. \ref{fig:sse}b exhibits, similar to YIG/Pt, a clear signal maximum near $T = \SI{120}{\kelvin}$, but with a significantly reduced signal amplitude.

\begin{figure}[bt]
	\centering
	\includegraphics[width=8.01 cm]{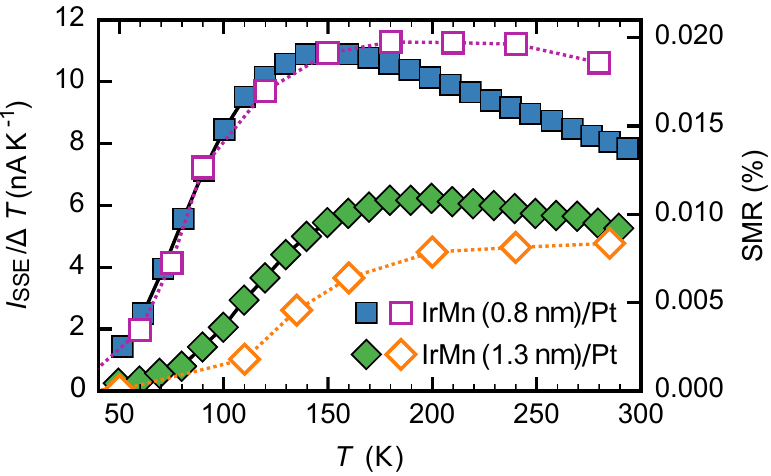}
	\caption{Comparison between temperature-dependent SSE (closed symbols) and SMR (open symbols) amplitudes for YIG/IrMn (\SI{0.8}{\nano\meter})/Pt (blue squares) and YIG/IrMn (\SI{1.3}{\nano\meter})/Pt (green diamonds).}
	\label{fig:fig3}
\end{figure}

Next, we compare the results of SSE measurements with the results of the SMR measurements to understand and differentiate between interface and spin transport effects.
The temperature-dependent SMR amplitudes obtained by the angular-dependent measurements are shown in Fig. \ref{fig:fig3} (open symbols), directly compared to the ISHE current amplitude (closed symbols).
Apart from a small difference in the amplitude ratio, both SMR and SSE feature similar temperature-dependent profiles with an overlapping, strong signal suppression that sets in at low temperatures.

In the following, we discuss the results above to understand the measured signals and the different contributions.
To deduce information about the spin current transmission details across IrMn, we analyze and compare the different data sets obtained for the different sample stacks individually:
Firstly, we discuss the temperature-dependent generation and detection of magnon spin currents.
For that we consider the bilayers of YIG/Pt and YIG/IrMn, which do not involve spin current transmission across the full IrMn layer.
In YIG/Pt, as shown in Fig. \ref{fig:sse}a, the detected spin Seebeck current exhibits a distinct amplitude maximum near $T = \SI{90}{\kelvin}$, which was explained before as a consequence of an increasing magnon propagation length in YIG with decreasing temperature, counteracted by a reduced occupation of magnon states due to lower thermal energy \cite{Guo2016}.
However, rather than being a pure bulk effect of the FM, the position of the signal maximum also depends on the employed ISHE detection layer \cite{Guo2016,cramer2017magnon}, implying a spectral-dependent transmission of magnons across the YIG/metal interface.
YIG/IrMn (Fig. \ref{fig:sse}b) shows a qualitatively similar behavior as compared to YIG/Pt but with a shifted peak position near $T = \SI{120}{\kelvin}$, which can be explained from the different magnon mode transmissions for YIG/Pt and YIG/IrMn as discussed for different detection layers in the literature \cite{Guo2016,cramer2017magnon}.

Next, we discuss the spin current transport and to understand its properties, we compare the stacks YIG/IrMn/Pt and YIG/IrMn.
The large difference in the SSE signal amplitude for YIG/IrMn and YIG/IrMn/Pt can be easily understood considering material properties such as a smaller spin Hall angle ($\theta_{\mathrm{SH}}^{\mathrm{IrMn}} \approx 0.8 \, \theta_{\mathrm{SH}}^{\mathrm{Pt}}$ \cite{Mendes2014}), a shorter spin diffusion length ($\lambda_{\mathrm{sf}}^{\mathrm{IrMn}} = \SI{0.7}{\nano\meter}$ vs. $\lambda_{\mathrm{sf}}^{\mathrm{Pt}} = \SI{2}{\nano\meter}$ \cite{Zhang2014,Isasa2015}) as well as a higher film resistivity ($\sigma^{\mathrm{IrMn}}/\sigma^{\mathrm{Pt}} \approx 0.15$ \cite{Mendes2014}) of IrMn as compared to Pt.
We now look closely at the comparison between YIG/IrMn (\SI{1.3}{\nano\meter}) (purple diamond, Fig. \ref{fig:sse}b) and YIG/IrMn (\SI{1.3}{\nano\meter})/Pt (green diamond, Fig. \ref{fig:sse}a).
Given the much lower signal amplitude of YIG/IrMn as compared to YIG/IrMn/Pt and furthermore the much lower resistance of the Pt, it is clear that in the YIG/IrMn/Pt sample the signal contribution from the ISHE voltage generation in the IrMn is negligible.
Thus, we can interpret the YIG/IrMn/Pt signal as the pure signal of the spin current transmitted from the YIG across the IrMn into the Pt, where due to the ISHE it is converted into the measured voltage.

Comparing the temperature dependences, we find in YIG/IrMn (\SI{1.3}{\nano\meter}) a clear signal maximum near $T = \SI{120}{\kelvin}$, while in YIG/IrMn (\SI{1.3}{\nano\meter})/Pt at temperatures below \SI{150}{\kelvin} the signal is strongly attenuated.
To explain this key feature of the strong attenuation, we go through all the processes to identify the origin: 
(i) We have established from the YIG/Pt system measurements that the spin current generated in the YIG and the detection in the Pt are large below \SI{150}{\kelvin} (Fig. \ref{fig:sse}a).
(ii) From the YIG/IrMn system, we know that the spin transport across the YIG/IrMn interface below \SI{150}{\kelvin} is large (Fig. \ref{fig:sse}b). 
Hence, what remains to explain the attenuation of the signal below \SI{150}{\kelvin} in the YIG/IrMn/Pt system is the spin transport across the IrMn, which apparently is suppressed below \SI{150}{\kelvin}.
The transmission of the spin current can be of both electronic and magnonic nature, with the temperature dependence of $I_{\mathrm{SSE}} / \Delta T$ in YIG/IrMn/Pt implying that the dominating contribution to the spin transport is strongly suppressed at low temperatures.

Hence, we need to understand whether the magnonic or the electronic spin current dominates.
From the fact that the signal in the YIG/IrMn system is still large below \SI{150}{\kelvin}, we deduce that the charge-based spin currents in the IrMn, which are necessary for the ISHE so they can be converted into a charge current signal, are also still large at temperatures below \SI{150}{\kelvin}. 
The observed strong attenuation of the measured signal in the YIG/IrMn/Pt system thus must stem from the magnonic spin current transport across the IrMn layer.
Finally and importantly this is then also in line with the theory put forward above, where a short spin transport length is found for antiferromagnetically ordered systems.

\begin{figure}[bt]
	\centering
	\includegraphics[width = 79.7 mm]{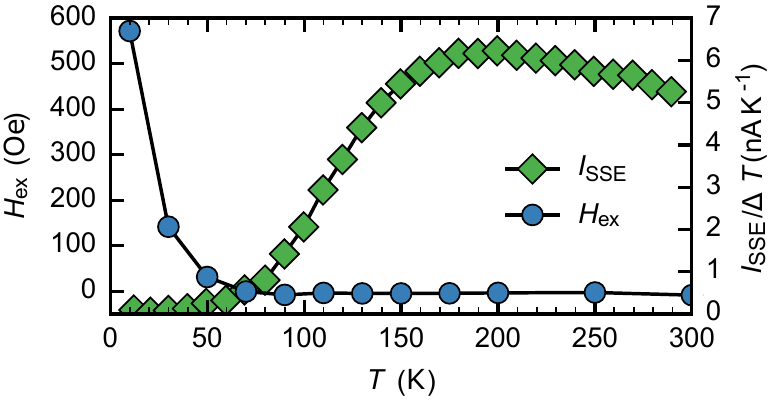}
	\caption{(a) Exchange-bias anisotropy field detected in SiO\textsubscript{2}/IrMn (\SI{1.3}{\nano\meter})/CoFe (\SI{2}{\nano\meter}) (blue circles) and (b) spin Seebeck current measured for YIG/IrMn (\SI{1.3}{\nano\meter})/Pt (green diamonds) as a function of temperature.}
	\label{fig:fig4}
\end{figure}

To further reinforce this interpretation of a potential relation of our experimental findings with the phase transition between the antiferromagnetic and the paramagnetic phase, we performed temperature-dependent magnetometry measurements on a SiO\textsubscript{2}/IrMn (\SI{1.3}{\nano\meter})/CoFe (\SI{2}{\nano\meter}) reference sample.
This reference sample is necessary to identify the transition temperature as the very large thickness of the used YIG films does not allow one to observe exchange-bias in the YIG/IrMn/Pt samples used for the transport experiments.
From the magnetometry data, the additional exchange anisotropy field of the IrMn film exerted on the CoFe layer is extracted as a function of temperature, see Fig. \ref{fig:fig4}.
The exchange-bias field vanishes at the so-called blocking temperature $T_B \approx \SI{80}{\kelvin}$, which in thin films usually is found to be smaller than $T_{\mathrm{N\acute{e}el}}$ \cite{van2000difference}.
While the absolute value needs to be taken with care, however, considering the compositional differences of the investigated samples, the $\mathrm{N\acute{e}el}$ temperature of the  YIG/IrMn (\SI{1.3}{\nano\meter})/Pt stack is expected to be below \SI{150}{\kelvin}.
One observes that above $T_B$, $I_{\mathrm{SSE}}$ starts to increase significantly in the corresponding sample, which we identify as a further indication for a correlation between the signal suppression and the AFM phase transition of the IrMn film.
Above the $\mathrm{N\acute{e}el}$ temperature, the magnonic spin current can be transported by short-range correlations \cite{chatterji2009antiferromagnetic}, while below $T_{\mathrm{N\acute{e}el}}$ the AFM magnon gap (see. Eq. \ref{afmgap}) in IrMn opens up and increases when further decreasing the temperature.
According to the physical processes depicted in Fig. \ref{rotevan}, this signifies a transition from spin angular momentum transfer via precessing spin waves to evanescent waves at low temperatures, which can explain the strong suppression of $I_{\mathrm{SSE}}/\Delta T$ due to the strong decay of the evanescent waves.

Therefore, from all the indications, we conclude that the spin current is at least partially transported by AFM magnonic spin currents in the IrMn layer.
This conclusion is further corroborated by recent studies by Saglam \textit{et al.} \cite{Saglam2016}, who report on two transport regimes in Ni\textsubscript{80}Fe\textsubscript{20}/FeMn/W systems with varying FeMn thickness. 
In the short-range regime (small thickness), spin propagation is dominated by electronic transport, whereas in the long-range regime (larger thickness) magnonic excitations yield the leading spin transport channel.
Note that FeMn exhibits a larger spin-diffusion length as IrMn \cite{Zhang2014}.
Furthermore, in the experiment by Saglam \textit{et al.} the spin current is emitted by the Ni\textsubscript{80}Fe\textsubscript{20} FMR mode excited at $f = \SI{9}{\giga\hertz}$, whereas in SSE experiments thermal magnons up to the THz regime are present.

The correlation between the AFM order in IrMn and its spin current propagation properties becomes furthermore apparent when considering the trilayer samples with varying IrMn thickness.
Whereas the thickness-dependent reduction of $I_{\mathrm{SSE}}/\Delta T$ is to be understood as a result of spin diffusion (either electronic and magnonic), the thickness-dependent critical temperature for signal suppression is a direct indication of the paramagnetic-antiferromagnetic phase transition.
In agreement with the findings by Frangou \textit{et al.} \cite{Frangou2016}, who report an increasing $T_{\mathrm{N\acute{e}el}}$ with increasing IrMn thickness, the signal suppression for thicker IrMn sets in at higher temperatures.

Finally, the comparison of SSE and SMR amplitudes reveals very good agreement (Fig. \ref{fig:fig3}), showing in particular coinciding low-temperature behavior, despite the conceptional differences of the underlying effects.
The SMR includes strong interface effects, considering that the pure spin current induced in a heavy metal due to the SHE interacts with the surface spins of an adjacent magnetic layer \cite{Nakayama2013}, which results in a spin-orientation-dependent film resistance.
The SSE, on the other hand, includes the conversion of bulk magnon spin currents into electronic spin currents and eventually charge currents by the ISHE.
Taking into account the differences of thickness, conductivity and spin Hall angle of Pt and IrMn, one can assume that in the SMR experiment the SHE spin current is mainly generated in the Pt layer.
The observed angular dependence of the resistance change corresponds to a \textit{positive} SMR that appears in systems in which the spin currents interact with the surface magnetization of FMs. 
For AFMs, on the other hand, the SMR follows the N$\mathrm{\acute{e}}$el order parameter and a \textit{negative} SMR is observed \cite{hou2017tunable,baldrati2017negative,fischer2017spin}.
Therefore, we conclude that for the SMR signal measured, the spin current that is generated in the Pt transmits across the IrMn and interacts with the YIG surface magnetization (absorption/reflection).
Potential \textit{negative} SMR contributions may appear at magnetic fields of sufficient strength to align and rotate the N$\mathrm{\acute{e}}$el order parameter in IrMn, which is not the case here.
Assuming the validity of the aforementioned magnonic spin transport mechanism in IrMn, the coinciding temperature dependences of SSE and SMR amplitudes imply a strong coupling of the electronic spin current in Pt to the order parameter in IrMn at the IrMn/Pt interface and a dominating contribution of the spin transport across the IrMn layer for the temperature dependence.

\section{Summary}
In conclusion, we have studied both theoretically and experimentally the propagation of pure spin currents in antiferromagnetic systems.
While in insulating AFMs spin information transmission is exclusively provided by magnonic excitations, metallic AFMs as well can exhibit charge-mediated spin currents.
AFM magnons exhibit a high-frequency gap. Despite the high velocity of antiferromagnetic magnons close to the frequency gap, the analytical model of magnonic transport shows that AFM magnons decay on much shorter distances, due to a shorter and frequency-independent lifetime.
Using atomistic spin dynamics simulations, we demonstrate the propagation of spin waves from a FM to an AFM and show that short range evanescent modes are excited below the frequency gap, whereas normal modes with a longer propagation length are excited above the frequency gap.
Beyond theoretical considerations, we furthermore investigate spin transmission across the metallic AFM IrMn by temperature-dependent SSE and SMR measurements in YIG/IrMn, YIG/Pt and YIG/IrMn/Pt heterostructures.
From a systematic comparison of the obtained results, we conclude that the spin currents are at least partially mediated by AFM magnons.
At low temperatures, where IrMn orders antiferromagnetically, the detected spin signals in YIG/IrMn/Pt transmitted across the IrMn become strongly suppressed, whereas in YIG/IrMn a notable signal induced by solely an electronic spin current is still detected.
This is explained by the AFM magnon gap in IrMn to open up, such that the spin current is transported by evanescent waves that exhibit a strong decay over the film thickness.
Furthermore, the critical temperature, at which the suppression sets in, increases with increasing IrMn thickness as expected for a thickness-dependent phase transition temperature.
Eventually, the coinciding temperature dependences observed for SSE and SMR suggest strong interaction of the electronic spin current in Pt towards the order parameter in the AFM IrMn.

\section*{Acknowledgements}
 The authors would like to thank the Deutsche Forschungsgemeinschaft (DFG) for financial support (SPP 1538 “Spin Caloric Transport", SFB767 in Konstanz and SFB TRR173 in Mainz), the Graduate School of Excellence Materials Science in Mainz (DFG/GSC 266), the EU projects (IFOX FP7-NMP3-LA-2012246102, INSPIN FP7-ICT-2013-X 612759), ERATO "Spin Quantum Rectification Project" (No. JPMJER1402) from JST, Japan, Grant-in-Aid for Scientific Research on Innovative Area "Nano Spin Conversion Science" (No. JP26103005) and Grant-in-Aid for young scientists (B) (No. JP17K14331) from JSPS KAKENHI, Japan. 
 
\section*{References}

\bibliographystyle{iopart-num}
\bibliography{./library2}

\end{document}